\documentclass{article}

\usepackage{arxiv}

\usepackage[utf8]{inputenc} 
\usepackage[T1]{fontenc}    
\usepackage{hyperref}       
\usepackage{url}            
\usepackage{booktabs}       
\usepackage{amsfonts}       
\usepackage{nicefrac}       
\usepackage{microtype}      
 \usepackage{longtable}
 \usepackage{multirow}
 \usepackage{graphics}

\title{A Review of Bayesian Statistical Approaches for Big Data}

\author{
 Farzana Jahan\thanks{Corresponding author}\\
  School of Mathematical Sciences\\
ARC Centre of Mathematical and Statistical Frontiers\\
  Science and Engineering Faculty\\
  Queensland University of Technology \\
  Brisbane, Queensland, Australia\\
  \texttt{f.jahan@hdr.qut.edu.au} \\
   \And
 Insha Ullah \\
 School of Mathematical Sciences\\
  ARC Centre of Mathematical and Statistical Frontiers\\
  Science and Engineering Faculty\\
  Queensland University of Technology \\
  Brisbane, Queensland, Australia\\
  \texttt{insha.ullah@qut.edu.au} \\
   \AND
    Kerrie L. Mengersen\\
  School of Mathematical Sciences\\
  ARC Centre of Mathematical and Statistical Frontiers\\
  Science and Engineering Faculty\\
  Queensland University of Technology \\
  Brisbane, Queensland, Australia\\
  \texttt{k.mengersen @qut.edu.au} \\
}

\begin{document}
\maketitle

\begin{abstract}
The modern era is characterised as an era of information or Big Data. This has motivated a huge literature on new methods for extracting information and insights from these data. A natural question is how these approaches differ from those that were available prior to the advent of Big Data. We present a review of published studies that present Bayesian statistical approaches specifically for Big Data and discuss the reported and perceived benefits of these approaches. We conclude by addressing the question of whether focusing only on improving computational algorithms and infrastructure will be enough to face the challenges of Big Data.
\end{abstract}


\section{Introduction}
\label{sec 1}
Although there are many variations on the definition of Big Data \cite{jifa2014data,de2015big,de2016formal,wamba2015big}, it is clear that it encompasses large and often diverse quantitative data obtained from increasing numerous sources at different individual, spatial and temporal scales, and with different levels of quality. Examples of Big Data include data generated from social media \cite{bello2016social}; data collected in biomedical and healthcare informatics research such as DNA sequences and electronic health records \cite{luo2016big}; geospatial data generated by remote sensing, laser scanning, mobile mapping, geo-located sensors, geo-tagged web contents, volunteered geographic information (VGI), global navigation satellite system (GNSS) tracking and so on \cite{li2016geospatial}. The volume and complexity of Big Data often exceeds the capability of the standard analytics tools (software, hardware, methods and algorithms) \cite{kaisler2013big,gandomi2015beyond}. The concomitant challenges of managing, modelling, analysing and interpreting these data have motivated a large literature on potential solutions from a range of domains including statistics, machine learning and computer science. This literature can be grouped into four broad categories of articles. The first includes general articles about the concept of Big Data, including the features and challenges, and their application and importance in specific fields. The second includes literature concentrating on infrastructure and management, including parallel computing and specialised software. The third focuses on statistical and machine learning models and algorithms for Big Data. The final category includes articles on the application of these new techniques to complex real-world problems. 

In this chapter, we classify the literature published on Big Data into finer classes than the four broad categories mentioned earlier and briefly reviewed the contents covered by those different categories. But the main focus of the chapter is around the third category, in particular on statistical contributions to Big Data. We examine the nature of these innovations and attempt to catalogue them as modelling, algorithmic or other contributions. We then drill further into this set and examine the more specific literature on Bayesian approaches. Although there is an increasing interest in this paradigm from a wide range of perspectives including statistics, machine learning, information science, computer science and the various application areas, to our knowledge there has not yet been a review of Bayesian statistical approaches for Big Data. This is the primary contribution of this chapter. 

This chapter provides a review of the published studies that present Bayesian statistical models specifically for Big Data and discusses the reported and perceived benefits of these approaches. We conclude by addressing the question of whether focusing only on improving computational algorithms and infrastructure will be enough to face the challenges of Big Data.

The chapter proceeds as follows. In the next section, literature search and inclusion criteria for this chapter is outlined. A classification of Big Data literature along with brief review of relevant literature in each class is presented in section 3. Section 4 consists of a brief review of articles discussing Big Data problems from statistical perspectives, followed by a review of Bayesian approaches applied to Big Data. The final section includes a discussion of this review with a view to answering the research question posed above.

\section{Literature Search and Inclusion Criteria}
\label{sec:2}
The literature search for this review paper was undertaken using different methods. The search methods implemented to find the relevant literature  and the criteria for the inclusion of the literature in this chapter are briefly discussed in this section.

\subsection{Inclusion Criteria}
Acknowledging the fact that there has been a wide range of literature on Big Data, the specific focus in this chapter was on recent developments published in the last 5 years, 2013-2019. 

For quality assurance reasons, of the literature only peer reviewed published articles, book chapters and conference proceedings were included in the chapter. Some articles were also included from arXiv and pre-print versions for those to be soon published and from well known researchers working in that particular area of interest.  

\subsection{Search Methods}
\textbf{Database Search:} The database ``Scopus" was used to initiate the literature search. To identify the availability of literature and broadly learn about the broad areas of concentration, the following keywords were used: Big Data, Big Data Analysis, Big Data Analytics, Statistics and Big Data.

The huge range of literature obtained by this initial search was complemented by a search of ``Google Scholar" using more specific key words as follows: Features and Challenges of Big Data, Big Data Infrastructure, Big Data and Machine Learning, Big Data and Cloud Computing, Statistical approaches/methods/models in Big Data, Bayesian Approaches/Methods/Models in Big Data, Big Data analysis using Bayesian Statistics, Bayesian Big Data, Bayesian Statistics and Big Data.

\textbf{Expert Knowledge:} In addition to the literature found by the above Database search, we used expert knowledge and opinions in the field and reviewed the works of well known researchers in the field of Bayesian Statistics for their research works related to Bayesian approaches to Big Data and included the relevant publications for review in this chapter. 

\textbf{Scanning References of selected literature:} Further studies and literature were found by searching the references of selected literature.

\textbf{Searching with specific keywords:} Since the focus of this chapter is to review the Bayesian approaches to Big Data,  more literature was sourced by using specific Bayesian methods or approaches found to be applied to Big Data: Approximate Bayesian Computation and Big Data, Bayesian Networks in Big Data, Classification and regression trees/Bayesian Additive regression trees in Big Data, Naive Bayes Classifiers and Big Data, Sequential Monte Carlo and Big Data, Hamiltonian Monte Carlo and Big Data, Variational Bayes and Big Data, Bayesian Empirical Likelihood and Big Data, Bayesian Spatial modelling and Big Data, Non parametric Bayes and Big Data.

This last step was conducted in order to ensure that this chapter covers the important and emerging areas of Bayesian Statistics and their application to Big Data. These searches were conducted in ``Google Scholar" and up to 30 pages of results were considered in order to find relevant literature.  

\section{Classification of Big Data literature}
\label{sec:3}
The published articles on Big Data can be divided into finer classes than the four main categories described above. Of course, there are many ways to make these delineations. Table 1 shows one such delineation, with representative references from the last five years of published literature. The aim of this table is to indicate the wide ranging literature on Big Data and provide relevant references in different categories for interested readers.  

\begin{figure}[h!]
\centering
  \includegraphics{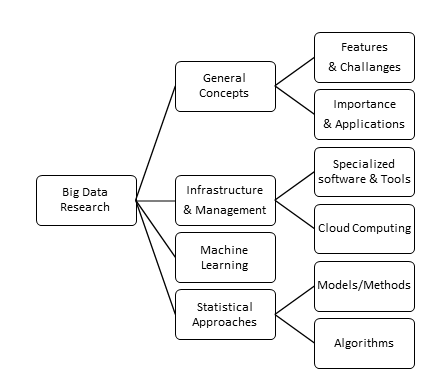}
  \caption{Classification of Big Data literature}
  \label{Figure 1}
\end{figure}

\begin {longtable}  { p{3.5 cm} p{8 cm}}
\caption {Classes of Big Data Literature} \\
\hline
Topic & Representative References \\
\hline
Features and Challenges & \cite{sivarajah2017critical,xia2017small,oprea2016big,de2015big,de2016formal,wamba2015big,gandomi2015beyond,emani2015understandable,fan2014challenges,sagiroglu2013big}.\\
\hline
Infrastructure & \cite{kousar2018multi,zhang2018convergence,magdon2017tpcx,siddiqa2017big,vyascomparative,apiletti2017frequent,zhang2016parallel,muller2016utilizing,oancea2014integrating,pandey2014prominence,watson2014tutorial,das2013big,liu2013computing}. \\
\hline
Cloud computing & \cite{manibharathi2018survey,moustafa2017collaborative,cai2017iot,yang2017big,assunccao2015big,loebbecke2015reflections,branch2014cloud,zhang2014hybrid,demirkan2013leveraging,o2013big,talia2013clouds}.\\
\hline
Applications (3 examples) & Social science: \cite{albury2017data,cappella2017vectors,bello2016social,mansour2016understanding,shah2015big,burrows2014after}.

Health/medicine/medical science: \cite{mahlmann2018big,cheung2018moving,alonso2017systematic,bansal2016big,rumsfeld2016big,alyass2015big,belle2015big,binder2015big,brennan2015nursing,viceconti2015big,bates2014big,raghupathi2014big,roski2014creating,yoo2014big,hay2013big}. 

Business: \cite{raguseo2018big,ducange2018glimpse,sun2018big,fosso2017big,bradlow2017role,akter2016big,bughin2016big,marshall2015leading}. \\
\hline
Machine Learning Methods & \cite{divya2018machine,bibault2016big,obermeyer2016predicting,akusok2015high,al2015efficient,landset2015survey,suthaharan2014big,bifet2014big,fahad2014survey,huang2014big}. \\
\hline
Statistical Methods & \cite{Dunson2018,nongxa2017mathematical,WebbVargas2017,Wang2017,Wang2017b,Liu2017,zhang2017exact,franke2016statistical,schifano2016online,hilbert2016big,Wang2016,doornik2015statistical,Pehlivanl2015,wise2015theory,Chen2015,Sysoev2014,chen2014stochastic,hoerl2014applying,hensman2013gaussian}.\\
\hline
Bayesian Methods & \cite{gunawan2018subsampling,quiroz2018speeding,zhu2017big,wu2017average,lee2018abc,minsker2017robust,lindsten2017divide,lee2016forest,liu2016stein,scott2016bayes,hassani2015forecasting,tripuraneni2015particle,moores2015pre,ma2014bayesian,strathmann2015unbiased,zoubin2013scaling,allenby2014perspectives,liu2013scalable}.\\
\hline
\end {longtable}

The links between these classes of literature can be visualised as in Figure 1 and a brief description of each of the classes and the contents covered by the relevant references listed are provided in Table 2. The brief reviews presented in Table 2 can be helpful for interested readers to develop a broad idea about each of the classes mentioned in Table 1. However, Table 2 does not include brief reviews of the last two classes, namely, Statistical Methods and Bayesian Methods, since these classes are discussed in detail in sections 4 and 5. We would like to acknowledge the fact that Bayesian methods are essentially part of statistical methods, but in this chapter, the distinct classes are made intentionally to be able to identify and discuss the specific developments in Bayesian approaches. 

\begin {longtable}  { p{11.5 cm}}
\caption {Brief Review of relevant literature under identified classes} \\
\hline
Features and Challenges \\
\hline \begin{itemize}
    \item The general features of Big Data are volume, variety, velocity, veracity, value  \cite{sagiroglu2013big, de2016formal}  and some salient features include massive sample sizes and high dimensionality \cite{sagiroglu2013big}.
    \item Many challenges of Big Data regarding storage, processing, analysis and privacy are identified in the literature \cite{de2016formal, oprea2016big,sagiroglu2013big,fan2014challenges,emani2015understandable}. 
\end{itemize}\\
\hline
Infrastructure \\
\hline \begin{itemize}
    \item To manage and analyse Big Data, infrastructural support is needed such as sufficient storage technologies and data management systems. These are being continuously developed and improved. MangoDB, Terrastore and RethhinkDb are some examples of the storage technologies; more on evolution technologies with their strengths, weaknesses, opportunities and threats are available in \cite{siddiqa2017big}.
    \item To analyse Big Data, parallel processing systems and scalable algorithms are needed. MapReduce is one of the pioneering data processing systems \cite{zhang2016parallel}. Some other useful and popular tools to handle Big Data are Apache, Hadoop, Spark \cite{apiletti2017frequent}.
\end{itemize}\\
\hline
Cloud computing \\
\hline \begin{itemize}
    \item Cloud computing, the practice of using a network of remote servers hosted on the Internet  rather than a local server or a personal computer, plays a key role in Big Data analysis by providing required infrastructure needed to store, analyse, visualise and model Big Data using scalable and adaptive systems \cite{assunccao2015big}. 
    \item  Opportunities and challenges of cloud computing technologies, future trends and application areas are widely discussed in the literature \cite{yang2017big,branch2014cloud,talia2013clouds} and new developments on cloud computing are proposed to overcome known challenges, such as collaborative anomaly detection \cite{moustafa2017collaborative}, hybrid approach for scalable sub-tree anonymisation using MapReduce on cloud \cite{zhang2014hybrid} etc. 
\end{itemize}\\
\hline
Applications (3 examples)\\
\hline \begin{itemize}
    \item Big Data has made it possible to analyse social behaviour and an individual's interactions with social systems based on social media usage \cite{albury2017data,shah2015big,burrows2014after}. Discussions on challenges and future of social science research using Big Data have been made in the literature \cite{shah2015big,cappella2017vectors}. 
    \item  Research involving Big Data in medicine, public health, biomedical and health informatics has increased exponentially over the last decade \cite{cheung2018moving,raghupathi2014big,roski2014creating,binder2015big,luo2016big,bates2014big}. Some examples include infectious disease research \cite{bansal2016big,hay2013big}, developing personalised medicine and health care \cite{viceconti2015big,alyass2015big} and improving cardiovascular care \cite{rumsfeld2016big}.
    \item Analysis of Big Data is used to solve many real world problems in business, in particular, using Big Data analytics for innovations in leading organisations \cite{marshall2015leading}, predictive analytics in retail \cite{bradlow2017role},  analysis of business risks and benefits \cite{raguseo2018big}, development of market strategies \cite{ducange2018glimpse} and so on. The opportunities and challenges of Big Data in e-commerce and Big Data integration in business processes can be found in the review articles by \cite{akter2016big} and \cite{wamba2015big}. 
\end{itemize}\\
\hline
Machine Learning Methods\\
\hline
 \begin{itemize}
    \item Machine learning is an interdisciplinary field of research primarily focusing on theory, performance, properties of learning systems and algorithms \cite{qiu2016survey}. Traditional machine learning is evolving  to tackle the additional challenges of Big Data \cite{qiu2016survey,al2015efficient}.
    \item Some examples of developments in machine learning theories and algorithms for Big Data include high performance machine learning toolbox \cite{akusok2015high}, scalable machine learning online services for Big Data real time analysis \cite{baldominos2014scalable}.
    \item There is a large and increasing research on specific applications of machine learning tools for Big Data in different disciplines. For example, \cite{obermeyer2016predicting} discussed the future of Big Data and machine learning in clinical medicine; \cite{azar2015dimensionality} discussed a classifier specifically for medical Big Data and \cite{bibault2016big} reviewed the state of art and future prospects of machine learning and Big Data in radiation oncology. 
\end{itemize}\\
\hline\\
\end {longtable}

\section{Statistical Approaches to Big Data}
\label{sec:4}

The importance of modelling and theoretical considerations for analysing Big Data are well stated in the literature \cite{hilbert2016big,wise2015theory}. These authors pointed out that blind trust in algorithms without proper theoretical considerations will not result in valid outputs. The emerging challenges of Big Data are beyond the issues of processing, storing and management. The choice of suitable statistical methods is crucial in order to make the most of the Big Data \cite{franke2016statistical, hoerl2014applying}. \cite{Dunson2018} highlighted the role of statistical methods for interpretability, uncertainty quantification, reducing selection bias in analysing Big Data. 

In this section we present a brief review of some of the published research on statistical perspectives, methods, models and algorithms that are targeted to Big Data. As above, the review is confined to the last five years, commencing with the most recent contributions. Bayesian approaches are reserved for the next section.
\newpage
\begin {longtable}{p{4.5in}}
\caption {Classification of statistical literature to Big Data}\\
\hline
Topic: Discussion Article\\
\hline
Author: Dunson (2018) \cite{Dunson2018}
\begin{itemize}
\item Discussed the background of big data from the perspectives of the machine learning and statistics communities.
\item Listed the differences in the methods and inferences as replicability, uncertainty quantification, sampling, selection bias and measurement error drawn from statistical perspectives to those of machine learning.
\item Identified the statistical challenges for high dimensional complex data (big data) in quantifying uncertainty, scaling up sampling methods and selection of priors in Bayesian methods.
\end{itemize}\\
\hline
Topic: Review\\
\hline
Author: Nongxa (2017) \cite{nongxa2017mathematical}
\begin{itemize} 
\item Identified challenges of big data as: high dimensionality, heterogeneity and incompleteness, scale, timeliness, security and privacy.
\item Pointed out that  mathematical and statistical challenges of big data require updating the core knowledge areas (i.e., linear algebra, multivariable calculus, elementary probability and statistics, coding or programming) to more advanced topics (i.e., randomised numerical linear algebra, topological data analysis, matrix and tensor decompositions, random graphs; random matrices and complex networks ) in mathematical and statistical education.
\end{itemize}\\
\hline
Author: Franke et al.(2016) \cite{franke2016statistical} 
\begin{itemize}
\item Reviewed different strategies of analysis as: data wrangling, visualisation, dimension reduction, sparsity regularisation, optimisation, measuring distance, representation learning, sequential learning and provided detailed examples of applications.
\end{itemize}\\
\hline
Author: Chen et al. (2015) \cite{Chen2015} 
\begin{itemize} 
\item Emphasised the importance of statistical knowledge and skills in Big Data Analytics using several examples.
\item Discussed some statistical methods that are useful in the context of big data as: confirmatory and exploratory data analysis tools, data mining methods including supervised learning (classification, regression/prediction) and unsupervised learning (cluster analysis, anomaly detection, association rule learning), visualisation techniques etc.
\item Elaborated on the computational skills needed for statisticians in data acquisition, data processing, data management and data analysis.
\end{itemize}\\
\hline
Author: Hoerl et al. (2014) \cite{hoerl2014applying}
\begin{itemize} 
\item Provided a background of big data reviewing relevant articles.
\item Discussed the importance of statistical thinking in big data problems reviewing some misleading results produced by sophisticated analysis of big data without involving statistical principles.
\item Elaborated on the roles of statistical thinking for data quality, domain knowledge, analysis strategies in order to solve complex unstructured problems involving big data.
\end{itemize}\\
\hline
Topic: Review of methods \& extension \\
\hline
Author: Wang et al. (2016) \cite{Wang2016} 
\begin{itemize} 
\item Reviewed statistical methods and software packages in R and recently developed tools to handle Big Data, focusing on three groups: sub-sampling, divide and conquer and online processing. 
\item Extended the online updating approach by employing variable selection criteria.
\end{itemize}\\
\hline
Topic: Methods review, new methods\\
\hline
Author: Genuer et al. (2017) \cite{Genuer2017} 
\begin{itemize} 
\item Reviewed proposals dealing with scaling random forests to big data problems.
\item Discussed subsampling, parallel implementations, online processing of random forests in detail.
\item Proposed five variants of Random Forests for big data.
\end{itemize}\\
\hline 
Author: Wang and Xu (2015) \cite{Wang2015} 
\begin{itemize} 
\item Reviewed different clustering methods applicable to big data situations.
\item Proposed a clustering procedure with adaptive density peak detection applying multivariate kernel estimation and demonstrated the performance through simulation studies and analysis of a few benchmark gene expression data sets. 
\item Developed a R-package “ADPclust” to implement the proposed methods.
\end{itemize}\\
\hline 
Author: Wang et al. (2017) \cite{Wang2017}
\begin{itemize} 
\item Proposed a method and algorithm for online updating implementing bias corrections with extensions for application in a generalised linear model (GLM) setting.
\item Evaluated the proposed strategies in comparison with previous algorithms \cite{schifano2016online}.
\end{itemize}\\
\hline 
Topic: New methods and algorithms\\
\hline
Author: Liu et al. (2017) \cite{Liu2017} 
\begin{itemize}
\item Proposed a novel sparse GLM with L0 approximation for feature selection and prediction in big omics data scenarios.
\item Provided novel algorithm  and software in MATLAB (L0ADRIDGE) for performing L0 penalised GLM in ultra high dimensional big data.
\item Comparison of performance with other methods (SCAD, MC+) using simulation and real data analysis (mRNA,microRNA, methylation data from TGCA ovarian cancer).
\end{itemize}\\
\hline
Author: Schifano et al. (2016) \cite{schifano2016online} 
\begin{itemize}
\item Developed new statistical methods and iterative algorithms for analysing streaming data.
\item Proposed methods to enable update of the estimations and models with the arrival of new data.
\end{itemize}\\
\hline
Author: Allen et al. (2014) \cite{Allen2014} 
\begin{itemize} 
\item Proposed generalisations to Principal Components Analysis (PCA)  to take into account structural relationships in big data settings.
\item Developed fast computational algorithms using the proposed methods (GPCA, sparse GPCA and functional GPCA) for massive data sets.
\end{itemize}\\
\hline
Topic: New algorithms\\
\hline
Author: Wang and Samworth (2017) \cite{Wang2017b} 
\begin{itemize}
\item Proposed a new algorithm "inspect" (informative sparse projection for estimation of change points) to estimate the number and location of change points in high dimensional time series.
\item The algorithm, starting from a simple time series model, was extended to detect multiple change points and was also extended to have spatial or temporal dependence, assessed using simulation studies and real data application.
\end{itemize}\\
\hline
Author: Yu and Lee (2017) \cite{Yu2017} 
\begin{itemize} 
\item Extended the alternating direction method of multipliers (ADMM)  to solve penalised quantile regression problems involving massive data sets having faster computation and no loss of estimation accuracy.
\end{itemize}\\
\hline
Author: Zhang and Yang (2017) \cite{zhang2017exact} 
\begin{itemize}
\item Proposed new algorithms using ridge regression to make it efficient for handling big data.
\end{itemize}\\
\hline
Author: Doonrik and Hendry (2015)  \cite{doornik2015statistical} 
\begin{itemize} 
\item Discussed the statistical model selection algorithm "autometrics" for econometric data \cite{doornik2009autometrics} with its application to fat big data (having larger number of variables than the number of observations) . 
\item Extended algorithms for tackling computational issues of fat big data applying block searches and re-selection by lasso for correlated regressors.
\end{itemize}\\
\hline
Author: Sysoev et al. (2014) \cite{Sysoev2014} 
\begin{itemize} 
\item Presented efficient algorithms to estimate bootstrap or jackknife type confidence intervals for  fitted big data sets by Multivariate Monotonic Regression.
\item Evaluated the performance of the proposed algorithms using a case study on death in coronary heart disease for a large population.
\end{itemize}\\
\hline
Author: Pehlivanl(2015) \cite{Pehlivanl2015}
\begin{itemize} 
\item Proposed a novel approach for feature selection from high dimensional data.
\item Tested the efficiency of the proposed method using sensitivity, specificity, accuracy and ROC curve.
\item Demonstrated the approach on micro-array data.
\end{itemize}\\
\hline
\end {longtable}

Among the brief reviews of the relevant literature in Table 3, we include detailed reviews of three papers which are more generic in explaining the role of statistics and statistical methods in Big Data along with recent developments in this area. 

\cite{Wang2016} summarised the published literature on recent methodological developments for Big Data in three broad groups: subsampling, which calculates a statistic in many subsamples taken from the data and then combining the results \cite{politis1999subsampling}; divide and conquer, the principle of which is to break a dataset into smaller subsets to analyse these in parallel and combine the results at the end \cite{srivastava2018scalable} ; and online updating of streaming data \cite{schifano2016online}, based on online recursive analytical processing. He summarised the following methods in the first two groups: subsampling based methods (bag of little bootstraps, leveraging, mean log likelihood, subsample based MCMC), divide and conquer (aggregated estimating equations, majority voting, screening with ultra high dimension, parallel MCMC). The authors, after reviewing existing online updating methods and algorithms, extended the online updating of stream data method by including criterion based variable selection with online updating. The authors also discussed the available software packages (open source R as well as commercial software) developed to handle computational complexity involving Big Data. For breaking the memory barrier using R, the authors cited and discussed several data management packages (sqldf, DBI, RSQLite, filehash, bigmemory, ff) and packages for numerical calculation (speedglm, biglm, biganalytics, ffbase, bigtabulate, bigalgebra, bigpca, bigrf, biglars, PopGenome). The R packages for breaking computing power were cited and discussed in two groups: packages for speeding up (compiler, inline, Rcpp, RcpEigen, RcppArmadilo, RInside, microbenchmark, proftools, aprof, lineprof, GUIprofiler) and packages for scaling up (Rmpi, snow, snowFT, snowfall, multicore, parallel, foreach, Rdsm, bigmemory, pdpMPI, pbdSLAP, pbdBASE, pbdMAT, pbdDEMO, Rhipe, segue, rhbase, rhdfs, rmr, plymr, ravroSparkR, pnmath, pnmath0, rsprng, rlecuyer, doRNG, gputools, bigvis). The authors also discussed the developments in Hadoop, Spark, OpenMP, API and using FORTRAN and C++ from R in order to create flexible programs for handling Big Data. The article also presented a brief summary about the commercial statistical software, e.g., SAS, SPSS, MATLAB. The study included a case study of fitting a  logistic model to a massive data set on airline on-time performance data from the 2009 ASA Data Expo mentioning the use of some R packages discussed earlier to handle the problem with memory and computational capacity. Overall, this study provided a comprehensive review and discussion of state-of-the-art statistical methodologies and software development for handling Big Data. 

\cite{Chen2015} presented their views on the challenges and importance of Big Data and explained the role of statistics in Big Data Analytics based on a review of relevant literature. This study emphasised the importance of statistical knowledge and skills in Big Data Analytics using several examples. As detailed in Table 3, the authors broadly discussed a range of statistical methods which can be really helpful in better analysis of Big Data, such as, the use of exploratory data analysis principle in Statistics to investigate correlations among the variables in the data or establish causal relationships between response and explanatory variables in the Big Data. The authors specifically mentioned hypothesis testing, predictive analysis using statistical models, statistical inference using uncertainty estimation to be some key tools to use in Big Data analysis. The authors also explained that the combination of statistical knowledge can be combined with the Data mining methods such as unsupervised learning (cluster analysis, Association rule learning, anomaly detection) and supervised learning (regression and classification) can be beneficial for Big Data analysis. The challenges for the statisticians in coping with Big Data were also described in this article, with particular emphasis on computational skills in data acquisition (knowledge of programming languages, knowledge of web and core communication protocols), data processing (skills to transform voice or image data to numeric data using appropriate software or programming), data management (knowledge about database management tools and technologies, such as NoSQL) and scalable computation (knowledge about parallel computing, which can be implemented using MapReduce, SQL etc.).

As indicated above, many of the papers provide a summary of the published literature which is not replicated here. Some of these reviews are based on large thematic programs that have been held on this topic. For example, the paper by \cite{franke2016statistical} is based on presentations and discussions held as part of the program on Statistical Inference, Learning and Models for Big Data which was held in Canada in 2015. The authors discussed the four V’s (volume, variety, veracity and velocity) of Big Data and mentioned some more challenges in Big Data analysis which are beyond the complexities associated with the four V’s. The additional ``V" mentioned in this article is veracity. Veracity refers to biases and noise in the data  which may be the result of the heterogeneous structure of the data sources, which may make the sample non representative of the population. Veracity in Big Data is often referred to as the biggest challenge compared with the other V's. The paper reviewed the common strategies for Big Data analysis starting from data wrangling which consists of data manipulation techniques for making the data eligible for analysis; visualisation which is often an important tool to understand the underlying patterns in the data and is the first formal step in data analysis; reducing the dimension of data using different algorithms such as Principal Component Analysis (PCA) to make Big Data models tractable and interpretable; making models more robust by enforcing sparsity in the model by the use of regularisation techniques such as variable selection and model fitting criteria; using optimisation methods based on different distance measures proposed for high dimensional data and by using different learning algorithms such as representation learning and sequential learning. Different applications of Big Data were shown in public health, health policy, law and order, education, mobile application security, image recognition and labelling, digital humanities and materials science.

There are few other research articles focused on statistical methods tailored to specific problems, which are not included in Table 2. For example, \cite{Castruccio2016} proposed a statistics-based algorithm using a stochastic space-time model with more than 1 billion data points to reproduce some features of a climate model. Similarly, \cite{McCormick2014} used various statistical methods to obtain associations between drug-outcome pairs in a very big longitudinal medical experimental database (with information on  millions of patients) with a detailed discussion on the big results problem by providing a comparison of statistical and machine learning approaches. Finally, \cite{hensman2013gaussian} proposed stochastic variational inference for Gaussian processes which makes the application of Gaussian process to huge data sets (having millions of data points).

From the review of some relevant literature related to statistical perspectives for analysing Big Data, it can be seen that along with scaling up existing algorithms, new methodological developments are also in progress in order to face the challenges associated with Big Data.

\section{Bayesian Approaches in Big Data}
\label{sec:5}

As described in the Introduction, the intention of this review is to commence with a broad scope of the literature on Big Data, then focus on statistical methods for Big Data, and finally to focus in particular on Bayesian approaches for modelling and analysis of Big Data. This section consists of a review of published literature on the last of these.

There are two defining features of Bayesian analysis: (i) the construction of the model and associated parameters and expectations of interest, and (ii) the development of an algorithm to obtain posterior estimates of these quantities. In the context of Big Data, the resultant models can become complex and suffer from issues such as unavailability of a likelihood, hierarchical instability, parameter explosion and identifiability. Similarly, the algorithms can suffer from too much or too little data given the model structure, as well as problems of scalability and cost.  These issues have motivated the development of new model structures, new methods that avoid the need for models, new Markov chain Monte Carlo (MCMC) sampling methods, and alternative algorithms and approximations that avoid these simulation-based approaches. We discuss some of the concomitant literature under two broad headings, namely computation and models realising that there is often overlap in cited papers.

\subsection{Bayesian Computation}
\label{subsec:1}
In Bayesian framework a main-stream computational tool has been the Markov chain Monte Carlo (MCMC). The traditional MCMC methods do not scale well because they need to iterate through the full data set at each iteration to evaluate the likelihood \cite{wu2017average}.  Recently several attempts have been made to scale MCMC methods up to massive data. A widely used strategy to overcome the computational cost is to distribute the computational burden across a number of machines. The strategy is generally referred to as divide-and-conquer sampling. This approach breaks a massive data set into a number of easier to handle subsets, obtains posterior samples based on each subset in parallel using multiple machines and finally combines the subset posterior inferences to obtain the full-posterior estimates \cite{srivastava2018scalable}. The core challenge is the recombination of sub-posterior samples to obtain true posterior samples. A number of attempts have been made to address this challenge. 

\cite{neiswanger2013asymptotically} and \cite{white2015piecewise} approximated the sub-posteriors using kernel density estimation and then aggregated the sub-posteriors by taking their product. Both algorithms provided consistent estimates of the posterior. \cite{neiswanger2013asymptotically} provided faster MCMC processing since it allowed the machine to process the parallel MCMC chains independently. However, one limitation of the asymptotically embarrassing parallel MCMC algorithm \cite{neiswanger2013asymptotically} is that it only works for real and unconstrained posterior values, so there is still scope of works to make the algorithm work under more general settings. 

\cite{wang2013parallelizing} adopted a similar approach of parallel MCMC but used a Weierstrass transform to approximate the sub-posterior densities instead of a kernel density estimate. This provided better approximation accuracy, chain mixing rate and potentially faster speed for large scale Bayesian analysis. 

\cite{scott2016bayes} partitioned the data at random and performed MCMC independently on each subset to draw samples from posterior given the data subset. To obtain consensus posteriors they proposed to average samples across subsets and showed the exactness of the algorithm under a Gaussian assumption. This algorithm is scalable to a very large number of machines and works in cluster, single multi core or multiprocessor computers or any arbitrary collection of computers linked by a high speed network. The key weakness of consesnsous MCMC is it does not apply to non Gaussian posterior. 

\cite{minsker2017robust} proposed dividing a large set of independent data into a number of non-overlapping subsets, making inferences on the subsets in parallel and then combining the inferences using the median of the subset posteriors. The median posterior (M-posterior) is constructed from the subset posteriors using Weiszfeld's algorithm, which provides a scalable algortihm for robust estimation . 

\cite{guhaniyogi2018meta} extended this notion to spatially dependent data, provided a scalable divide and conquer algorithm to analyse big spatial data sets named spatial meta kriging. The multivariate extension of spatial meta kriging has been addressed by \cite{guhaniyogi2018multivariate}. These approaches of meta kriging are practical developments for Bayesian spatial inference for Big Data, specifically with ``big-N" problems \cite{lasinio2013discussing}.    

\cite{wu2017average} proposed a new and flexible divide and conquer framework by using re-scaled sub-posteriors to approximate the overall posterior. Unlike other parallel approaches of MCMC, this method creates artificial data for each subset, and applies the overall priors on the artificial data sets to get the subset posteriors. The sub-posteriors are then re-centred to their common mean and then averaged to approximate the overall posterior. The authors claimed this method to have statistical justification as well as mathematical validity along with sharing same computational cost with other classical parallel MCMC approaches such as consensus Monte Carlo, Weierstrass sampler. \cite{bouchard2018bouncy} proposed a non-reversible rejection-free MCMC method, which reportedly outperforms state-of-the-art methods such as: HMC, Firefly by having faster mixing rate and lower variances for the estimators for high dimensional models and large data sets. However, the automation of this method is still a challenge. 

Another strategy for scalable Bayesian inference is the sub-sampling based approach. 
In this approach, a smaller subset of data is queried in the MCMC algorithm to evaluate the likelihood at every iteration.  
\cite{maclaurin2014firefly} proposed to use an auxiliary variable MCMC algorithm that evaluates the likelihood based on a small subset of the data at each iteration yet simulates from the exact posterior distribution.
To improve the mixing speed,  \cite{korattikara2014austerity} used an approximate Metropolis Hastings (MH) test based on a subset of data. A similar approach is used in \cite{bardenet2014towards}, where the accept/reject step of MH evaluates the likelihood of a random subset of the data. \cite{bardenet2017markov} extended this approach by replacing a number of likelihood evaluations by  a Taylor expansion centred at the maximum of the likelihood and concluded that their method outperforms the previous algorithms \cite{korattikara2014austerity}.

The scalable MCMC approach was also improved by \cite{quiroz2015scalable} using a difference estimator to estimate the log of the likelihood accurately using only a small fraction of the data. \cite{quiroz2018speeding} introduced an unbiased estimator of the log likelihood based on weighted sub-sample which is used in the MH acceptance step in speeding up based on a weighted MCMC efficiently. Another scalable adaptation of MH algorithm was proposed by \cite{maire2017informed} to speed up Bayesian inference in Big Data namely informed subsampling MCMC which involves drawing of subsets according to a similarity measure (i.e., squared L2 distance between full data and maximum likelihood estimators of subsample) instead of using uniform distribution. The algorithm showed excellent performance in the case of a limited computational budget by approximating the posterior for a tall dataset.

Another variation of MCMC in Big Data has been made by \cite{strathmann2015unbiased}. These authors approximated the posterior expectation by a novel Bayesian inference framework for approximating the posterior expectation from a different perspective suitable for Big Data problems, which involves paths of partial posteriors. This is a parallelisable method which can easily be implemented using existing MCMC techniques. It does not require the simulation from full posterior, thus bypassing the complex convergence issues of kernel approximation. However, there is still scope for future work to look at computation-variance trade off and finite time bias produced by MCMC.

Hamiltonian Monte Carlo (HMC) sampling methods provide powerful and efficient algorithms for MCMC using high acceptance probabilities for distant proposals \cite{chen2014stochastic}. A conceptual introduction to HMC is presented by \cite{betancourt2017conceptual}. \cite{chen2014stochastic} proposed a stochastic gradient HMC using second-order Langevin dynamics. Stochastic Gradient Langevin Dynamics (SGLD) have been proposed as a useful method for applying MCMC to Big Data where the accept-reject step is skipped and decreasing step size sequences are used \cite{ahn2014distributed}. For more detailed and rigorous mathematical framework, algorithms and recommendations, interested readers are referred to \cite{teh2016consistency}. 

A popular method of scaling Bayesian inference, particularly in the case of analytically intractable distributions, is Sequential Monte Carlo (SMC) or particle filters \cite{chopin2013smc2, beskos2015sequential, gunawan2018subsampling}. SMC algorithms have recently become popular as a method to approximate integrals. The reasons behind their popularity include their easy implementation and parallelisation ability, much needed characteristics in Big Data implementations \cite{lee2016forest}. SMC can approximate a sequence of probability distributions on a sequence of spaces with an increasing dimension by applying resampling, propagation and weighting starting with the prior and eventually reaching to the posterior of interest of the cloud of particles. \cite{gunawan2018subsampling} proposed a sub-sampling SMC which is suitable for parallel computation in Big Data analysis, comprising two steps. First, the speed of the SMC is increased by using an unbiased and efficient estimator of the likelihood, followed by a Metropolis within Gibbs kernel. The kernel is updated by a HMC method for model parameters and a block-pseudo marginal proposal for the auxiliary variables \cite{gunawan2018subsampling}. Some novel approaches of SMC include: divide-and-conquer SMC \cite{lindsten2017divide}, multilevel SMC \cite{beskos2015sequential}, online SMC \cite{gloaguen2018online} and one pass SMC \cite{lin2013online}, among others. 

Stochastic variational inference (VI, also called Variational Bayes, VB) is a faster alternative to MCMC \cite{hoffman2013stochastic}. It approximates probability densities using a deterministic optimisation method \cite{liu2016stein} and has seen widespread use to approximate posterior densities for Bayesian models in large-scale problems. The interested reader is referred to \cite{blei2017variational} for a detailed introduction to variational inference designed for statisticians, with applications. VI has been implemented in scaling up algorithms for Big Data. For example, a novel re-parameterisation of VI has been implemented for scaling latent variable models and sparse GP regression to Big Data \cite{gal2014distributed}.

There have been studies which combined the VI and SMC in order to take advantage from both strategies in finding the true posterior \cite{donnet2017shortened,naesseth2017variational,rabinovich2015variational}.  \cite{naesseth2017variational} employed a SMC approach to get an improved variational approximation, \cite{rabinovich2015variational}  by splitting the data into block, applied SMC to compute partial posterior for each block and used a variational argument to get a proxy for the true posterior by the product of the partial posteriors. 
The combination of these two techniques in a Big Data context was made by \cite{donnet2017shortened}. \cite{donnet2017shortened} proposed a new sampling scheme called Shortened Bridge Sampler, which combines the strength of deterministic approximations of the posterior that is variational Bayes with those of SMC. This sampler resulted in reduced computational time for Big Data with huge numbers of parameters, such as data from genomics or network.

\cite{guhaniyogi2014bayesian} proposed a novel algorithm for Bayesian inference in the context of massive online streaming data, extending the Gibbs sampling mechanism for drawing samples from conditional distributions conditioned on sequential point estimates of other parameters. The authors compared the performance of this conditional density filtering algorithm in approximating the true posterior with SMC and VB, and reported good performance and strong convergence of the proposed algorithm.

Approximate Bayesian computation (ABC) is gaining popularity for statistical inference with high dimensional data and computationally intensive models where the likelihood is intractable \cite{mckinley2018approximate}. A detailed overview of ABC can be found in \cite{Sisson2018overview} and asymptotic properties of ABC are explored in \cite{Frazier2018asymptotic}. ABC is a likelihood free method that approximates the posterior distribution utilising imperfect matching of summary statistics \cite{Sisson2018overview}.  Improvements on existing ABC methods for efficient estimation of posterior density with Big Data (complex and high dimensional data with costly simulations) have been proposed by \cite{izbicki2019abc}. The choice of summary statistics from high dimensional data is a topic of active discussion; see, for example, \cite{izbicki2019abc,singh2018multi}.
\cite{pudlo2015reliable} provided a reliable and robust method of model selection in ABC employing random forests which was shown to have a gain in computational efficiency. 

There is another aspect of ABC recently in terms of approximating the likelihood using Bayesian Synthetic likelihood or empirical likelihood \cite{drovandi2018approximating}. Bayesian 
synthetic likelihood arguably provides computationally efficient approximations of the likelihood with high dimensional summary statistics \cite{Meeds2014gps,wilkinson2014accelerating}. Empirical likelihood, on the other hand is a non-parametric technique of approximating the likelihood empirically from the data considering the moment constraints; this has been suggested in the context of ABC \cite{mengersen2013bayesian}, but has not been widely adopted. For further reading on empirical likelihood, see \cite{owen2001empirical}. 

Classification and regression trees are also very useful tools in data mining and Big Data analysis \cite{breiman2017classification}. There are Bayesian versions of regression trees such as Bayesian Additive Regression Trees (BART) \cite{chipman2010bart,kapelner2013bartmachine,allenby2014perspectives}. The BART algorithm has also been applied to the Big Data context and sparse variable selection by \cite{rockova2017posterior,van2017bayesian,linero2018bayesian}.

 Some other recommendations to speed up computations are to use graphics processing units \cite{lee2010utility,suchard2010understanding} and parallel programming approaches \cite{guha2012large,chang2013parallel,williamson2013parallel,ge2015distributed}.
 
\subsection{Bayesian Modelling}
\label{subsec:2}
The extensive development of Bayesian computational solutions has opened the door to further developments in Bayesian modelling. Many of these new methods are set in the context of application areas. For example, there have been applications of ABC for Big Data in many different fields \cite{Dutta2017,lee2018abc}. For example, \cite{Dutta2017} developed a high performance computing ABC approach for estimation of parameters in platelets deposition, while \cite{lee2018abc} proposed ABC methods for inference in high dimensional multivariate spatial data from a large number of locations with a particular focus on model selection for application to spatial extremes analysis. Bayesian mixtures are a popular modelling tool. VB and ABC techniques have been used for fitting Bayesian mixture models to Big Data \cite{mcgrory2007variational,tank2015streaming,hoffman2013stochastic,blei2017variational,moores2015pre}.

Variable selection in Big Data (wide in particular, having massive number of variables) is a demanding problem. \cite{liquet2017bayesian} proposed  multivariate extensions of the Bayesian group lasso for variable selection in high dimensional data using Bayesian hierarchical models utilising spike and slab priors with application to gene expression data. The variable selection problem can also be solved employing ABC type algorithms. \cite{Liu2018ABC} proposed a sampling technique, ABC Bayesian forests, based on splitting the data, useful for high dimensional wide data, which turns out to be a robust method in identifying variables with larger marginal inclusion probability. 
 
Bayesian non-parametrics \cite{muller2015bayesian} have unbounded capacity to adjust unseen data through activating additional parameters that were inactive before the emergence of new data. In other words, the new data are allowed to speak for themselves in non-parametric models rather than imposing an arguably restricted model (that was learned on an available data) to accommodate new data. The inherent flexibility of these models to adjust with new data by adapting in complexity makes them more suitable for Big Data as compared to their parametric counterparts. For a brief introduction to Bayesian non-parametric models and a nontechnical overview of some of the main tools in the area, the interested reader is referred to \cite{ghahramani2013bayesian}. 

The popular tools in Bayesian non-parametrics include Gaussian processes (GP) \cite{rasmussen2004gaussian}, Dirichlet processes (DP) \cite{rasmussen2000infinite}, Indian buffet process (IBP) \cite{ghahramani2006infinite} and infinite hidden Markov models (iHMM) \cite{beal2002infinite}. GP have been used for a variety of applications \cite{chalupka2013framework,damianou2013deep,buettner2015computational} and attempts have been made to scale it to Big Data \cite{hensman2013gaussian,hensman2015scalable,tran2015variational,deisenroth2015distributed}. 
DP have seen successes in clustering and faster computational algorithms are being adopted to scale them to Big Data \cite{wang2011online,wang2011fast,lin2013online, ma2014bayesian,ge2015distributed}. IBP are used for latent feature modeling, where the number of features are determined in a data-driven fashion and have been scaled to Big Data through variational inference algorithms \cite{zoubin2013scaling}. Being an alternative to classical HMM, one of the distinctive properties of iHMM is that it infers the number of hidden states in the system from the available data and has been scaled to Big Data using particle filtering algorithms \cite{tripuraneni2015particle}. 

Gaussian Processes are also employed in the analysis of high dimensional spatially dependent data \cite{banerjee2017high}. \cite{banerjee2017high} provided model-based solutions employing low rank GP and nearest neighbour GP (NNGP) as scalable priors in a hierarchical framework to render full Bayesian inference for big spatial or spatio temporal data sets. \cite{zhang2019practical} extended the applicability of NNGP for inference of latent spatially dependent processes by developing a conjugate latent NNGP model as a practical alternative to onerous Bayesian computations. Use of variational optimisation with structured Bayesian GP latent variable model to analyse spatially dependent data is made in \cite{atkinson2019structured}. For a review of methods of analysis of massive spatially dependent data including the Bayesian approaches, see \cite{heaton2017methods}.

Another Bayesian modelling approach that has been used for big and complex data is Bayesian Networks (BN). This methodology has generated a substantial literature examining theoretical, methodological and computational approaches, as well as applications \cite{tang2016bayesian}. BN belong to the family of probabilistic graphical models and based on direct acyclic graphs which are very useful representation of causal relationship among variables \cite{ben2008bayesian}. BN are used as efficient learning tool in Big Data analysis integrated with scalable algorithms \cite{wang2014scalable,zhou2017machine}. For a more detailed understanding of BN learning from Big Data, please see \cite{tang2016bayesian}. 

Classification is also an important tool for extracting information from Big Data and Bayesian classifiers, including Naive Bayes classifier (NBC) are used in Big Data classification problems \cite{katkar2013novel,liu2013scalable}. Parallel implementation of NBC has been proposed by \cite{katkar2013novel}. Moreover, \cite{liu2013scalable} evaluated the scalability of NBC in Big Data with application to sentiment classification of millions of movie review and found NBC to have improved accuracy in Big Data. \cite{ni2019scalable}  proposed a scalable multi step clustering and classification algorithm using Bayesian nonparametrics for Big Data with large n and small p which can also run in parallel. 

The past fifteen years has also seen an increase in interest in Empirical Likelihood (EL) for Bayesian modelling. The idea of replacing the likelihood  with an empirical analogue in a Bayesian framework was first explored in detail by \cite{lazar2003Bayesian}. The author demonstrated that this Bayesian Empirical Likelihood (BEL) approach increases the flexibility of EL approach by examining the length and coverage of BEL intervals. The paper tested the methods using simulated data sets. Later, \cite{schennach2005bayesian} provided probabilistic interpretations of BEL exploring moment condition models with EL and provided a non parametric version of BEL, namely Bayesian Exponentially Tilted Empirical Likelihood (BETEL). The BEL methods have been applied in spatial data analysis in \cite{chaudhuri2011empirical} and \cite{porter2015bayesian,porter2015multivariate} for small area estimation.   

We acknowledge that there are many more studies on the application of Bayesian approaches in different fields of interest which are not included in this review. There are also other review papers on overlapping and closely related topics. For example, \cite{zhu2017big} describes Bayesian methods of machine learning and includes some of the Bayesian inference techniques reviewed in the present study. However, the scope and focus of this review is different from that of \cite{zhu2017big}, which was focused around the methods applicable to machine learning. 

\section{Conclusions}
\label{sec:6}

We are living in the era of Big Data and continuous research is in progress to make most use of the available information. The current chapter has attempted to review the recent developments made in Bayesian statistical approaches for handling Big Data along with a general overview and classification of the Big Data literature with brief review in last 5 years. This review chapter provides relevant references in Big Data categorised in finer classes, a brief description of statistical contributions to the field and a more detailed discussion of the Bayesian approaches developed and applied in the context of Big Data.

On the basis of the reviews made above, it is clear that there has been a huge amount of work on issues related to cloud computing, analytics infrastructure and so on. However, the amount of research conducted from statistical perspectives is also notable. In the last five years, there has been an exponential increase in published studies focused on developing new statistical methods and algorithms, as well as scaling existing methods. These have been summarised in Section 4, with particular focus on Bayesian approaches in Section 5. In some instances citations are made outside of the specific period (see section 2) to refer the origin of the methods which are currently being applied or extended in Big Data scenarios.

With the advent of computational infrastructure and advances in programming and software, Bayesian approaches are no longer considered as being very computationally expensive and onerous to execute for large volumes of data, that is Big Data. Traditional Bayesian methods are now becoming much more scalable due to the advent of parallelisation of MCMC algorithms, divide and conquer and/or sub-sampling methods in MCMC, and advances in approximations such as HMC, SMC, ABC, VB and so on. With the increasing volume of data, non-parametric Bayesian methods are also gaining in popularity.

This review chapter aimed to review a range of methodological and computational advancement made in Bayesian Statistics for handling the difficulties arose by the advent of Big Data. By not focusing to any particular application, this chapter provided the readers with a general overview of the developments of Bayesian methodologies and computational algorithms for handling these issues. The review has revealed that most of the advancements in Bayesian Statistics for Big Data have been around computational time and scalability of particular algorithms, concentrating on estimating the posterior by adopting different techniques. However the developments of Bayesian methods and models for Big Data in the recent literature cannot be overlooked. There are still many open problems for further research in the context of Big Data and Bayesian approaches, as highlighted in this chapter.  

Based on the above discussion and the accompanying review presented in this chapter, it is apparent that to address the challenges of Big Data along with the strength of Bayesian statistics, research on both algorithms and models are essential.

\bibliographystyle{vancouver}  
\bibliography{arxiv_article} 

\section*{Acknowledgement}
This research was supported by an ARC Australian Laureate Fellowship for project, Bayesian Learning for Decision Making in the Big Data Era under Grant no. FL150100150. The authors also acknowledge the support of the Australian Research Council (ARC) Centre of Excellence for Mathematical and Statistical Frontiers (ACEMS). 

\section*{Appendix}
\addcontentsline{toc}{section}{Appendix}
List of acronyms used for Bayesian Computational Algorithms and Models are:
\begin {longtable}  { p{3.5 cm} p{9 cm}}
ABC & Approximate Bayesian Computation \\
BEL & Bayesian Empirical Likelihood\\
BN & Bayesian Network\\
BNN & Bayesian Neural Network\\
BART & Bayesian Additive Regression Trees\\
CART & Classification and Regression Trees\\
DP & Dirichlet Process\\
GP & Gaussian Process\\
HMC & Hamiltonian Monte Carlo\\
HMM & Hidden Markov Models \\
IBP & Indian Buffet Process\\
iHMM & Infinite Hidden Markov Models\\
MCMC & Markov Chain Monte Carlo\\
MH & Metropolis Hasting\\
NBC & Naive Bayes Classifier\\
NNGP & Nearest Neighbour Gaussian Process\\
SMC & Sequential Monte Carlo\\
VB & Variational Bayes\\
VI & Variational Inference\\
\end{longtable}

\end{document}